\documentclass{article}
\usepackage[utf8]{inputenc}
\usepackage[backend=biber]{biblatex}
\usepackage{url}
\usepackage{xcolor}
\usepackage{graphicx}
\usepackage{enumitem}
\usepackage{array}
\usepackage{ragged2e}
\usepackage{amsmath}
\usepackage{appendix}
\usepackage{authblk}
\usepackage{mathtools}
\usepackage{mathptmx}
\usepackage{mathrsfs}
\usepackage{amssymb}
\usepackage{amsmath}
\usepackage{amssymb}
\usepackage{csquotes}
\usepackage{hyperref}
\begin{document}
\title{Preventing or Mitigating Adversarial Supply Chain Attacks; a legal analysis}
\author[1]{Kaspar Rosager Ludvigsen}
\author[2]{Shishir Nagaraja} 
\author[3]{Angela Daly}
\affil[1]{Department of Computer and Information Sciences, University of Strathclyde, kaspar.rosager-ludvigsen@strath.ac.uk}
\affil[2]{Department of Computer and Information Sciences, University of Strathclyde, shishir.nagaraja@strath.ac.uk}
\affil[3]{Leverhulme Research Centre for Forensic Science and Dundee Law School, adaly001@dundee.ac.uk}

\begin{abstract}

The world is currently strongly connected through both the internet at large, but also the very supply chains which provide everything from food to infrastructure and technology. The supply chains are themselves vulnerable to adversarial attacks, both in a digital and physical sense, which can disrupt or at worst destroy them. In this paper, we take a look at two examples of such successful attacks and consider what their consequences may be going forward, and analyse how EU and national law can prevent these attacks or otherwise punish companies which do not try to mitigate them at all possible costs. We find that the current types of national regulation are not technology specific enough, and cannot force or otherwise mandate the correct parties who could play the biggest role in preventing supply chain attacks to do everything in their power to mitigate them. But, current EU law is on the right path, and further vigilance may be what is necessary to consider these large threats, as national law tends to fail at properly regulating companies when it comes to cybersecurity.
    
\end{abstract}


\maketitle

\section{Introduction}


Before the advent of widely adopted digital infrastructure systems, the biggest threats of information being stolen and by other means compromised was through the actions of its employees and from outside forces like spies and other intruders. By now, information as well as decisions can be altered, and even physical manifestations can be seen from these infiltration attempts and successes. Stuxnet\footnote{Attack occurred in 2010, see \url{https://www.wired.com/images_blogs/threatlevel/2010/11/w32_stuxnet_dossier.pdf}, last accessed 5 August 2022.} is an important and well known milestone that included physical consequences. It succeeded because it executed a serious of steps and actions and affected a monumental amount of physical and digital systems\footnote{Sometimes defences against these are secured by projects and principles created a long time ago too, see \url{https://krebsonsecurity.com/2022/05/when-your-smart-id-card-reader-comes-with-malware/}, last accessed 5 August 2022.}. This kind of stepping stone approach where one intrusion gives the attacker(s) access to an entire system of systems \parencite{roy_survey_2022}, is akin to an attack on an entire supply chain or system. This will be the greatest threat to any infrastructure supported by computers, civilian, commercial or military, and many do seem to care \parencite{creazza_who_2022}. Because of the potential consequences if a provider of security of a supply chain, or the parties of the supply chains themselves are compromised, we are interested in uncovering the means which these companies can be held accountable.




A variety of research has been done on Supply Chain Attacks on a organisational, supply chain and security level, but current legal measures are not well explored in the literature, with some exceptions \parencite{haber_cybersecurity_2017}. We will therefore comparatively analyse selected measures in several countries and the EU, both because of the diversity, but also because it gives a broad perspective and idea about how far we may be from properly regulating the risk of supply chain attacks, because studying cybersecurity subjects is without a doubt complicated and interdisciplinary \parencite{choo_interdisciplinary_2020, suryotrisongko_review_2019}.




Before this, we show two practical examples of Supply Chain Attacks. The first occurred in 2020 without physical consequences to the company SolarWinds Inc., which provides systems for managing software and other products. One difference between it and Stuxnet was that this affected everything from public authorities, including foreign intelligence of several countries, to wealthy private companies. As of the time of writing, we are not entirely sure what the purpose or the gains of the attack was, other than the US Department of Justice Confirming the compromise of their mailing environment\footnote{See \url{https://www.justice.gov/opcl/department-justice-statement-solarwinds-update}, last accessed 5 August 2022.}. 
Essentially, a sophisticated attack which enabled a multiplication of threats due to its system-wide access, which allowed it to act vertically and horizontally.



The other example is the Kaseya Ransomware Attack\footnote{\url{https://www.zdnet.com/article/kaseya-ransomware-attack-what-we-know-now/}, last accessed 5 August 2022.}, where another type of management structure was compromised and used as a means to inject ransomware into the users of it. Unlike the SolarWinds Inc. Attack, it was not caused by an equally sophisticated payload, but by a vulnerability that was discovered earlier. Its effect on commercial and public enterprises, as well as its clear physical consequence by striking down card payment systems in 100s of physical stores\footnote{\url{https://www.svt.se/nyheter/inrikes/it-attacken-mot-coop-detta-har-hant}, last accessed 5 August 2022.} makes it worth considering as an alternative to sophisticated attacks, which can cause the same type of damage. 

The article is structured in the following manner: Section 2 discusses issues with understanding and perceiving supply chains, and Section 3 elaborates on how supply chain attacks can be understood and defined, contains the two aforementioned examples, and a small suggestion how to improve and add to existing frameworks and understandings of the concept. We follow this in Section 4 with an analysis of the current legal rules concerning the needed security within the EU and selected national law, and future problems which will arise with the increased use of IoT and CPS which apply to supply chain attacks. In Section 5 we discuss future work, and finally we add some concluding remarks in Section 6.

We find that the two examples of Supply Chain Attacks illustrate two types of threats, either sophisticated and complicated attacks akin to Stuxnet, or extremely simple abuse of exploits. This shows that simple attacks such as those found in Social Engineering \parencite{holt_social_2020} are an equally big threat to expensively built custom malware by state actors. We also note that any future understandings of Supply Chain Attacks, be it on digital or physical supply chains, should use existing ideas of empirically recording and understanding threats, but should equally be open to anything from everywhere, since this is the security reality everyone faces, with the motivation of the adversary being paramount to not misunderstand. Adversaries may always have further interests than just breaching or stealing information. In the process of investigating whether the EU or the UK, Denmark or Ireland had any specific legislation built to prevent Supply Chain Attacks, we found none in a literal sense, but we listed relevant cybersecurity legislation and the legal means to punish companies which potentially did not prevent world-wide failures of digital or physical supply chains because of adversarial attacks. While not anywhere near enough, these legal means are real. However, they all rely on guidance or soft law to define the technology or the concept of supply chains, which we find to be lacking and risky, just like it is a problem in other product or sector specific legislation. Finally, we criticise the use of merely using fines, as this will not deter large companies, and suggest measures such as forceful closure or prison sentences for directors or staff for how severe the impact of failing to mitigate or prevent Supply Chain Attacks is.



\section{Supply Chains}

We note that the cybersecurity of digital as well as physical supply chains are characterised by:

\begin{enumerate}
    \item Excessive market tipping and monopolies.
    \item Network effects.
    \item Durability impacting complexity arising from software inter dependencies.
    \item Disintermediation of alternatives.
    \item Lack of transparency. 
\end{enumerate}

First point is the notion of increased use of specific operating systems, security systems or other specialised software or service, which then leads to monopolies or oligopolies in certain fields \parencite{economides_restrictions_2019}, benefiting no one but the providers. We refer to \parencite{duan_monopolies_2020} for a historical but crucial perspective, and see Section 3.1.1 for more on this issue with the first example. 

Second point stipulates the influence that products which have high user bases may cause, while the third shows the great weakness which the first two points cause. The complexity, or the lack thereof depending on the supply chain, will change and potentially lead to further weaknesses, and no standards or cybersecurity rules currently account for this. These usually rely on the corporate side to pick the right options or collaborations, but these may end up negatively impacting the durability and the safety of the supply chain if unsuitable.

The fourth point again relies on poor competition legislation and competition in practice. If there is no way to easily explain and understand what software and service solutions might be the best for a specific supply chain, myriads of vulnerabilities and failures become close to inevitable. This plays into issues in both competition as well as agreement and purchase law in national jurisdictions.

Finally, the fifth point relates to the issue of the lack of transparency. Public accountability and clear evidence for auditing and future lawsuits should demand this, both \textit{ex ante} and \textit{ex post} for when the failures do occur.  

These factors lead to a highly oligopolised landscape, which impacts how they can be controlled or not. Earlier regulation of fields like encryption have not gone well \parencite{ellis_history_1999, hellegren_history_2017, anderson_review_2021, pisaric_communications_2022}, and the sheer complexity of the many types of law affecting this subject go beyond this paper.

\section{Supply Chain Attacks}

In 2011, in a report made on behalf of Microsoft \parencite[P. 10 - 16]{charney_cyber_2011}, four key areas were identified and deemed to be important to cyber supply chain management for states; \textit{risk-based approach, transparency, flexibility and reciprocity.} Managing security, hardware and production chains is part of the first area, but it speaks against harshly legislating, and this is followed in the rest as well. Clearly, Microsoft wanted to indicate that government intervention should be kept at a minimal to keep their corporate influence high. Outside of this moot point, it shows that the debate was an equally high level more than ten years ago, and no other research indicates that the threat has lessened since then.

Supply Chain Attacks are adversarial attacks on a supply chain. The term is well defined because of its critical role, and has been discussed in detail in US government reports \parencite{miller_supply_2013, reed_supply_2014}. 
These are not academic papers, which creates a need for scrutiny of the models proposed. Both of these suggest a framework to understand the patterns and the structural dangers that these attacks pose, and they do it on the basis of attacks that did occur, see the appendixes in the reports for patterns. While this is sound, it does not leave room for anything which is not foreseen. 

The essential assumption for Supply Chain Attacks is that they can occur in all levels of the supply chain \parencite[P. 7]{miller_supply_2013}. From subcontractor to software or hardware development, to the highest primary party, anyone is a valid target. 

Technological development and the mindset of the attacker has however changed drastically since 2013. \textit{Miller} rightfully does not go into the risks proprietary software or hardware poses, but this still plays a role \parencite{zhou_impact_2022}. Because these tools are not developed by anyone in the supply chain, but can be attacked regardless of who uses it and who developed it, they must be a separate point to include. Malicious insertions are stated as the primary adversarial attack used against the supply chain, and they will often be multi-staged. This is still very much true, but we now have a range of attacks that do not involve insertions at any point. This could be subversion of control of CPS, leading to the destruction of goods or injuries \parencite{bonaci_make_2015}. For an overview of an area, see \parencite{hei_teleoperated_2022}. There would be no insertion into the software or hardware, but instead a manipulation of the communication channel to force the CPS to commit to orders not given by the user. 

Other authors do divide the attacks into categories. \textit{Eggers} writes that Supply Chain Attacks depend on the area which is targeted. Theft of IP, malicious substitution, alterations, malicious insertion, tampering and manipulation are just some of the many types than can occur \parencite[P. 886]{eggers_novel_2021}, but they can classically be viewed as falling under the CIA triad. In this case, loss of confidentiality would cover the theft of IP, while loss of integrity would occur during substitution and alterations, and tampering and manipulation would cause real loss of availability. Areas such as nuclear infrastructure which her paper concerns \parencite{eggers_nuclear_2020} clearly require very high degrees of caution.

Following this, we must change the current assumptions concerning Supply Chain Attacks to the following, to include every aspect and generalise it:

\begin{enumerate}
    \item Supply Chain Attacks can occur anywhere in the supply chain, and to any hardware or software in it, regardless of origin.
    \item The attacks can be of any kind. 
    \item The goal of the attacks must be more than to breach a given system. 
\end{enumerate}

What makes this different from a single adversarial attack on one device or system is that the aim is more than just initiation. The failure achieved by the system is therefore both the loss of for example integrity, but also the following loss of availability through ransomware or loss of confidentiality through privacy failures. We cannot quantify the goals of the attacker under most circumstances, because the perpetrators very rarely are identified, but we can derive them from their actions. Prevention or mitigation techniques include anything traditionally used against adversarial attacks, such as organisational measures, encryption and other classic measures. Of particular interest is mitigation at scale and through simulations and modelling \parencite{zheng_interdiction_2019, zheng_robust_2019}.





\subsection{Selected Examples}

We will in this section take a closer look at two recent supply chain attacks. One targeted all types of sectors, while the other was more focused on commercial targets. 

\subsubsection{Sunburst Backdoor}\label{solarwinds}

The first to discover this adversarial failure was the company FireEye, who in their report from 13 December 2020\footnote{\url{https://www.fireeye.com/blog/threat-research/2020/12/evasive-attacker-leverages-solarwinds-supply-chain-compromises-with-sunburst-backdoor.html}, last accessed 5 August 2022.} outline what their concerns are\footnote{A more detailed diagram of adversarial actions with Sunburst can be found here \url{https://www.fireeye.com/blog/threat-research/2020/12/sunburst-additional-technical-details.html}.}. The start of the attack was an update of the Orion IT monitoring and management software. Instead of a valid update, the users downloaded a trojan, and this occurred multiple times between March and May in 2020. What characterizes a trojan is its deceptive nature, with the original reference to the wooden horse used by the Greek Army in the \textit{Aeneid} by Virgil against the Trojans, to leave and hide soldiers inside, describing its purpose precisely. It included legitimate files except for one, the SolarWinds.Orion.Core.BusinessLayer.dll component, a dynamic-linked library file. These cannot be used on their own, and must be called up to have any function. This file would then be actively used by the legitimate Solarwinds.BusinessLayer executable file after a two week delay to enable the Sunburst backdoor. 


Before making contact back to the adversary, the trojan checks for anti-virus and other countermeasures, and a range of information about the machine that it is on. A very peculiar detail in this, is that it wants to avoid certain environments that are likely inside of SolarWinds Inc., and if it identifies that it is there, it will exit and cease to function after erasing its presence. This shows how specific the attack was, and how much the adversary wanted to avoid detection, but it was identified 7 months after its first entry into a client system by FireEye. The trojan mimicks natural SolarWinds API communication, which then enables it to connect to a domain that is controlled by the adversary, a so called command and control domain (C2). The trojan then tries to determine which security software resides on the hardware it is currently placed in, which it does locally and with great efficiency\footnote{For a list of all the types of software it would recognize, which is quite extensive, see \url{https://github.com/fireeye/sunburst_countermeasures/blob/main/fnv1a_xor_hashes.txt}, last accessed 5 August 2022.}. Even if it finds any of these, it will not exit because of it, instead checking for whether they are active, and whenever they are not, the trojan will disable the security software on the next power cycle in the Windows registry which it creates access to. When the trojan sees that none of the services on the list are active, because it has disabled them, it will initiate and let the adversary control it through the C2 domain. This is where the trojan can lead to a range of outcomes, with the most common being Teardrop. Sunburst is known to have dropped other payloads than Teardrop, which by itself is intriguing. Teardrop is purely a means to an end, through an extensive extraction process, including pretending to read information from a picture file, to drop a customized Cobalt Strike Beacon. The latter is modified proprietary software, defined as a asynchronous post-exploitation agent, which is usually used for penetration testing, but in this case has been directly used to attack a system. The beacon enables a massive amount of possible attacks. And with that, the backdoor enables for an adversary to do pretty much anything within the system. What the Sunburst Backdoor does and how it functions is without a doubt not novel, with possible links to existing malware being possible\footnote{See \url{https://securelist.com/sunburst-backdoor-kazuar/99981/}, last accessed 5 August 2022.}. This may put its sophistication into perspective \parencite{marelli_solarwinds_2022}.

\subsubsection{Kaseya Ransomware Attack}

Unlike the Sunburst Backdoor, this attack was a simpler process\footnote{\url{https://www.riskbasedsecurity.com/2021/07/12/the-kaseya-attack-everything-to-know/}, last accessed 5 August 2022.}. First, the attackers compromised the company Kaseya's Virtual Systems Administrator, with an exploit which was discovered some days prior\footnote{\url{https://csirt.divd.nl/2021/07/04/Kaseya-Case-Update-2/}, last accessed 5 August 2022.}. The program itself was only used in a limited amount of businesses, but most of those that ran it administered other companies' systems at the same time. Because of that, the compromise was exponentially increased by the nature of the service supply chain which the adversaries targeted. The adversaries used this to load ransomware onto a massive amount of businesses, including 800 Swedish Coop stores\footnote{\url{https://www.svt.se/nyheter/inrikes/it-attacken-mot-coop-detta-har-hant}, last accessed 5 August 2022.}. This is therefore a case of a service supply chain being compromised and used to target physical goods and service supply chains with a physical presence and product, and therefore a good example of a simple but effective Supply Chain Attack.

\subsection{Cyberphysical systems and IoT}

Physical and even digital supply chains have evolved since 2011. But CPS and IoT have dominated the world and especially the world of supply chains. In turn, this also affects which consequences Supply Chain Attacks can have on its targets. CPS refer to systems that have network access and which seamlessly integrate computation and physical components into operation \parencite{nsf_cyber-physical_2014}, and which usually have more than two levels, with sensors on the bottom, a network for these, and a top which controls the entire system \parencite{kobara_cyber_2016}. On the other side of this, we have the increased use of IoT, which act as network connected sensors that may part of a CPS or greater systems \parencite{xenofontos_consumer_2022}. The key between each is the network access, essentially a means to integrate a ``computer'' into anything, anywhere. An attack with simplicity of the one done on Kaseya can at any point knock out payment systems or physical stores, ticket dispensers or anything else that is loosely connected to a service supply chain above it. If the stores ran a system that was only controlled by the company itself, and no one else, this would not be possible, and general adversarial attacks would apply instead. 

\subsubsection{Security and Safety Constraints}

Increased use of systems of systems like CPS will therefore in turn decrease safety and decrease potential security. The first is due to all the ways these systems can fail. Any modern production facility will make use of CPS at IoT at once. For some meta commentary on this, see \parencite{nguyen_industrial_2019, tucker_sustainable_2021, de_las_morenas_security_2020}. This means that any attack on the main control systems will be able to shut down lower levels of the plant with ease, which in turn can cause a failure, either halting production or harming the employees. The same use of these systems decrease security overall, because the amount of entry points increase incrementally with the added features of IoT devices, each being an new door for adversaries \parencite{lagreca_survey_2017, wang_insecurity_2021}. 
IoT has a further issue, which is planned obsolescence. Unless produced and serviced by its users, IoT products have short lifespans \parencite{yousefnezhad_security_2020}, and after this they are to be considered significant security threats. If they are then a part of a greater CPS structure, they potentially risk loosing integrity, availability or integrity of the entire system.





\section{Law and Guidance}

Luckily, engineers seem to have affected some lawmakers and legislators, as there are special considerations taken in regards to supply chains and therefore also Supply Chain Attacks. But these are purely limited to guidance. Additionally, voluntary relationships between states and companies responsible for supply chains or cybersecurity exist, but cannot replace the needs for possible hard legal responses to attacks and failures on whole systems \parencite{shackelford_bottoms_2016, gyenes_voluntary_2014}. In this section, we will take a close look at relevant European Legislation, both in the Union as well as two Member States and the UK. 

\subsection{European Law}

The EU can only control certain areas because they are limited by competence. But they have provided the world of security with a great array of guidance as well as some legislation that can prove vital in the future.

\subsubsection{Security Legislation}

The most well known and used ``security legislation'', legislation that directly attempts to impose security obligations, in the European Union, is the NIS Directive\footnote{Directive 2016/1168, concerning measures for a high common level of security of network and information systems
across the Union.}. It is a directive and therefore requires implementation\footnote{For progress of the implementation, see \url{https://digital-strategy.ec.europa.eu/en/policies/nis-transposition}, last accessed 5 August 2022.} in each European Member State, which means there will be some divergence and legal fragmentation across the Union. 

Before we explain how this is relevant to the security of supply chains, we must justify whether it can be applied to them or not. This is done in national law (as we will shortly see below) through implementation, but if the directive itself is weak and does not lay out strong rules to be followed later on, it may be a pointless exercise. The NIS directive does not literally mention Supply Chain Attacks or adversarial attacks, but it is still relevant, because it supposedly sets up the infrastructure for the protection against them. For the Directive to apply, the supply chain must contain companies or public entities that are 'operators of essential services' defined in Article 4(4) and defined by the Member State in Article 5(2). Article 4(4) requires that they furthermore work within  Annex II, which has 7 broad categories, being \textit{energy, transport, banking, financial market infrastructures, drinking water supply and distribution, and digital infrastructure.} This leaves out providers of the security of these infrastructures, so cases like the Kaseya Rasomware Attack and the Sunburst Backdoor would not be covered. Firstly, because both of these companies are based in the US, and secondly because they are not included in any of these categories. However, it may possible to include them in an expanded version of the seventh category, but this is only doable through national law.

The first six categories are to be literally understood, but digital infrastructure needs an elaboration. The Directive sets out to cover IXPs, DNS service providers and TLD name registries. But there is nothing in the Directive that does not allow a Member State to include many more companies into their definition of ``operator of essential services''. An expanded definition of this could therefore be ISPs, SoMe providers, major security providers and more, and this would allow any Member State to force the NIS Directive to apply to those that are often responsible for the mitigation of Supply Chain Attacks. We will note whether any of the two Member States or the UK have done so in their implementation of the Directive.

Certain details of the implementation are done directly through the Commission Regulation\footnote{This is akin to legally binding guidance for the member states, not for anyone else, and is \textit{not} a normal EU regulation.} 2018/151, and while it does go more into detail when it comes to the security infrastructure required, it still refers to technical specifications elsewhere, not in the EU-law by itself. This document is used in national implementation.

The second piece of security legislation in the EU we will focus partially on\footnote{This act deserves its own paper for further security analysis, but it is relevant to discuss which influence it has the practical and real measures to mitigate Supply Chain Attacks.}, is the Cybersecurity Act\footnote{Regulation 2019/881}. Its title is deceptive, as it instead expands the powers of The European Union Agency for Cybersecurity (ENISA) and the initial process of cybersecurity certification\footnote{See additional analysis by \parencite{kamara_misaligned_2021, casarosa_cybersecurity_2022}.}. The Act has no literal details on Supply Chain Attacks or adversarial attacks in general, and its ideas concerning certification are not very promising to force security providers or others to prevent these. Initially, ENISA does not gain any powers that would transform it into a regulatory authority, it instead keeps its position as an advisory and guiding institution\footnote{See Art 3 and 4 of the Act.}. There is therefore no central overarching and controlling ``big brother'' when it comes to the regulation of security in the EU. There are however national regulators, but they are quite limited as to when they can enforce compliance. For the national authorities, the only times they can do act, is to withdraw certification from legal or physical entities regarding their software\footnote{See Art 56(8).}. They are not capable of anything else in a direct and effective sense\footnote{This depends on whether one views certification as an effective measure to increase security and prevent Supply Chain Attacks, or whether one prefers hard legal remedies and obligations.}. Because of this, we will not comment further on any practical or national consequences regarding the mitigation of Supply Chain Attacks by the Cybersecurity Act, and also because the certification scheme is (yet) not implemented or relevant on a European level. 


However, as indicated by the Act, ENISA publishes guidance and opinions and is supposed to be the central knowledge facilitator regarding security, and we will therefore go through some that are highly relevant to Supply Chain Attacks.
 
\subsubsection{Guidance}

To support the role of Supply Chain Attacks in the regime of the Directive, ENISA frequently publishes a threat landscape, and have their own taxonomy for the Supply Chain Attacks, which is less abstract and highly practical\footnote{\url{https://www.enisa.europa.eu/publications/threat-landscape-for-supply-chain-attacks}, last accessed 5 August 2022.}. Like the reports from the US, they are based on empirical information, in this case from incident reporting across the EU. Most of the content is therefore related to practical considerations and types of attacks, but they do include a list of recommendations. As much as these are interesting, they include references to fulfilling ISO and other standards, Google's End-to-End Framework for Supply Chain Integrity\footnote{\url{https://security.googleblog.com/2021/06/introducing-slsa-end-to-end-framework.html}, last accessed on 5 August 2022.} and other government recommendations\footnote{Like one written for the US government, which is generic and yet recommended by ENISA, see \url{https://d3fend.mitre.org/}, \textit{A knowledge graph of cybersecurity countermeasures}, last accessed 5 August 2022.}, and while these may be adequate, they are not in the spirit of security. Most of the technology and abstract ideas and security concepts that enable defences against Supply Chain Attacks are developed by academic researchers or other individuals \parencite{coleman_coding_2013}, and it would suit ENISA to follow suit and use more time developing the technical standards of their own, without reinventing the wheel, instead of just referring to existing ones. This may happen with the certification structure from the Cybersecurity Act, but the issue is still that no one dares setting hard technical standards or expectations for the security providers, leading to another ``hidden'' Wild West.

Regardless, this guidance makes a very important point that we need to keep in mind, that not everything is a Supply Chain Attack\footnote{P. 26.}. It can appear to be so, but it may be caused by design deficiencies or unpredictable behaviour of the software\footnote{Which can be made public but not used at any point by adversaries as well}, or it may simply be an adversarial attack that does not target links of the supply chain. A special methodological limitation is further added, in which a Supply Chain Attack that succeeds to infiltrate for example a service supply chain, like management software, but has targeted outdated versions of the software where users are not paying or part of the chain that the original manufacturer controls, will \textit{not} be considered a Supply Chain Attack.

\subsubsection{Other}

Other legislation will have security requirements tagged on through wording or through guidance. GDPR is an example of the first, product legislation like the MDR\footnote{Regulation (EU) 2017/745 of the European Parliament and of the Council of 5 April 2017 on medical devices, amending Directive 2001/83/EC, Regulation (EC) No 178/2002 and Regulation (EC) No 1223/2009 and repealing Council Directives 90/385/EEC and 93/42/EEC.} is the latter. The first works with the term ``state of the art''\footnote{Preamble 83.}, which refers to security, and therefore has vague requirements that are at least supposed to prevent abuse or leakage of personal data, but not mitigation of the Supply Chain Attack explicitly. Like any product legislation that includes digital infrastructures, the MDR has guidance issued by its central authority that should be followed. There is not legal requirement\footnote{This is up for debate currently.}, but it is heavily encouraged or even forced if caught before certification and release of the device\footnote{See 'Guidance on Qualification and Classification of Software in Regulation (EU) 2017/745 – MDR and Regulation (EU) 2017/746 – IVDR', \url{https://ec.europa.eu/docsroom/documents/37581}, last accessed 5 August 2022.}.

\subsection{National Law}


Unlike the overarching guidance and general rules of the EU, national law applies and functions directly onto the supply chains and its links. We here take a look at three different legal systems, and focus on how they each handle the threat (if at all)\footnote{Users are worth studying too, see, e.g., \parencite{ameen_keeping_2021} in this context.} and which other measures they provide to force manufacturers and other parties to mitigate these attacks as much as possible. The latter is not speculative, but a matter of showing the existing ways which security guidance or rules can be enforced, and considerations on expropriation or similar, even if the measures seem extreme or are close to impossible. Besides that, all three examples have also implemented the NIS directive. 

Manufacturers are legal entities, and each national state has rules to punish or otherwise force legal entities to comply, and even extreme measures such as expropriation if need be. Furthermore, national states can always act as private partners, and create contracts and arbitration systems that can further convince manufacturers and other parties to mitigate as many Supply Chain Attacks as possible. We therefore might not need to look to future for means and tools that can be used to increase security and safety for everyone. But there are emerging consequences from not applying national law to global private entities in this area \parencite{kilovaty_privatized_2019}.

All three countries share two measures that they can each implement directly. First, contractually binding providers of security and other supply chain parties, private to private party. Second, creating binding legal obligations for the supply chain at large, either specific security links or the main responsible parties\footnote{Or any combination of this.}. States can naturally be contract partners, and it is via these that they would be able to bind and force mitigation of Supply Chain Attacks through. The issue with doing so, is effectiveness and willingness of the participants. Any link of any major supply chain, or a security provider, has no interest in legally binding itself to terms without something in return, and solving issues in courts will as always be lengthy and costly. Arbitration clauses would be a possibility with such agreements, but since the state would act as a private party under those circumstances, any other actor could simply refuse to sign the contract in the first place. Let it be clear, that any measure from the state to force links of a supply chain to sign the contract, would constitute legal means, which would change the state from a private party to a public party in the contract, turning the legal relationship into something entirely different not unlike indirect regulation. This contract solution would be widely different between the UK and Ireland and Denmark respectively, due to the various roles of background law\footnote{Like case law on how contracts are viewed in Common Law versus how they are viewed in Scandinavian law where contract legislation plays a bigger role, although the latter now exists everywhere.}.

Creating new legal obligations is not novel, but currently none of our examples have direct legally binding obligations for the mitigation of Supply Chain Attacks. Any state, including or three examples, have the means to create these and enforce them as well, even if most supply chains are global or at least regional in their nature. Further limitations to this would be to prevent the circumvention by changing flags for shipping companies or head offices for the rest. Despite these measures existing, national states still have the power over individual workers or the physical infrastructure, therefore eliminating any arguments against the futility of the action. But making binding obligations will not make anyone popular.

\subsubsection{Denmark}

\paragraph{Implementation}
The NIS directive is implemented in Denmark via a range of laws and binding guidance\footnote{The main implementation act can be found here: \url{https://www.retsinformation.dk/eli/lta/2018/436}, last accessed 5 August 2022. The rest can be found under 'Yderligere dokumenter', then 'Se detaljeret overblik'.}. We will take a closer look at those that relate digital supply chains, but it is worth noting that there are strict direct requirements to levels of security for all 7 points mentioned in Annex II in the Directive, and each has its own binding guidance that the area \textit{must} follow or face fines\footnote{See for example security guidance for the electricty and gas providers, \url{https://www.retsinformation.dk/eli/lta/2021/2647}, last accessed 5 August 2022.}. The main implementation law defines essential services the same way as the Directive, but outside of essential financial service providers\footnote{\url{https://www.finanstilsynet.dk/tilsyn/information-om-udvalgte-tilsynsomraader/it-tilsyn/udpegelse_af_operatoerer_af_vaesentlige_tjenester}, last accessed 5 August 2022.} the exact list is secret or implied. § 4 is however the security specification, in that providers of essential services must control known risks, have adequate security compared to the risks and mitigate or prevent adversarial events from occurring to their systems\footnote{For perspectives outside of law, see \parencite{boeke_national_2018}.}.

Each different piece of guidance derived from the main implementation text may have different authorities being responsible. Fines are loosely defined and far lower than those in the other two examples, and this is due to a different culture regarding trust and a much tighter grip on public essential services. The latter enable changes and internal punishments for individuals based on labour law, not considered in the implementation, and restructuring or changes that could increase security without it being public or regulated tightly by the implementation law.

\paragraph{Other Measures}

The law of stock and partial companies in Danish law\footnote{Law nr. 763 of 23 July 2019.} enables the Danish Business Authority to forcefully close the most common types of companies in the country\footnote{See § 225, part one.}. The first two categories could theoretically be used for the failure to mitigate Supply Chain Attacks, but it is very unlikely, as forceful closure is usually related to rules of process or violation of minority shareholder or creditor rights. But, working against the purpose of the company as well as the ``wrong'' leadership are legitimate reasons, which is why it must be mentioned.

The other direct means which the Danish state has, is the expropriation of the company or the entire Supply Chain. This can theoretically be done on via the Danish Constitution\footnote{§ 73, part one, requires expanded view of 'property', which is acceptable since ownership of shares etc. is considered ``property'' of the individual, and companies are considered legal individuals owned and run by citizens.}, but has never been done in this way before. In a given situation where it would be necessary, such as during a national crisis, a freer and less restraining measure could be used instead, like contractual obligation or emergency obligations issued via law\footnote{This was seen on a widespread level during the Covid-19 pandemic, but this has so far not been regarding security and safety of supply chains.}. But its potential use would be all assets regarding the company, and would allow security of the Supply Chain if necessary. 

\subsubsection{United Kingdom} 

\paragraph{Implementation}

Even if the UK is an EU-member no longer, it implemented the NIS-directive when it entered into force\footnote{For other perspectives, see \parencite{carr_uk_2018}.}. It did so through different means than Denmark. The legal implementation is done through the Network and Information Systems Regulations 2018\footnote{Statute No. 506, 2018.} which designates competent authorities as those that can enforce the rules, and defines which types of penalties that are supposed to encourage compliance. Furthermore, the UK implemented a series of thorough guidance and systems, such as the Cyber Assessment Framework. The relevant authority, which depend on the area of essential services, has quite the range of powers, including right to retrieve information or inspect\footnote{See Part 5, 15.}, and penalties are fines\footnote{See 18(6).}. What is very intriguing are the grounds for the fines, which is either non-compliance through notices or not following orders, or not reporting incidents in various ways\footnote{See 17(10).}. Like Denmark, there seems to be no expansion of the concept of critical infrastructure to include security providers at large.   

\paragraph{Other Measures}

The UK can intervene and forcefully close companies and other legal entities. Unlike Denmark, the rules regarding this are tightly defined and leave little room for possible cases where security or mitigation of Supply Chain Attacks could be the basis of it. If the company is clearly defunct, which in some situations where destitute software is used may be the case, the company can be stricken off within 2 months\footnote{See the Companies Act 2006, 1000(3), assuming no answer is given from the company in question.}.

There is a theoretical possibility for something else in the Insolvency Act 1986, s 124 A. This section allows for winding up on grounds of public interest, which could include failure to comply with security requirements to prevent Supply Chain Attacks in the future. At present, the closest we get is closure due to fraud investigations, s 124 A, c, but because of how the section is shaped, it would be possible to add further reasons for winding up that could function as deterrence and reasons to comply. The case law concerning the statute further allows for closures within even more subjective terms\footnote{See, e.g., \textit{Re Alpha Club (UK) Ltd (2002)}, para 19.}.

Expanding the idea of expropriation in UK law to include punishments for damaging supply chains is difficult. Since there is no written constitution\footnote{See, e.g., \parencite{scarman_human_2012, frosini_is_2019}.}, we must rely on statutory law\footnote{Such as the Planning and Compulsory Purchase Act 2004.}, which is too specific and not reliant on case law to contain rights for the state that could include situations where a company and its assets must be acquired to mitigate Supply Chain Attacks. In terms of the other solutions and measures, the UK is therefore quite limited. 

\subsubsection{Ireland} 

\paragraph{Implementation}


Initially, Ireland has implemented the NIS directive through a Statutory Instrument like the UK, No. 360 of 2018, but its content and structure is quite different. As is the lack of deliberately abstract guidance which, like Denmark, does not exist. Definitions of operators of essential services and what is otherwise needed are here, but one noticeable difference is clear, as fines and investigations are done through either designated authorities\footnote{Statutory Instrument No. 360 of 2018, regulation(reg.) 7 and 8.} or authorized officers\footnote{Ibid, reg. 28.}. The latter is interesting, but does not mean there will be differences in enforcement, which is found in Reg 34. Like the other two jurisdictions, fines are the chosen tool, and they too have not expanded their concepts to include security providers at large.

\paragraph{Other Measures}

Rights and obligations of Companies and related authorities are regulated in the Companies Act 2014. Companies can be stricken off the register by the Registrar\footnote{Companies Act 2014, section 725.}, in our case if they fulfill the requirements set out in section 726. However, none of these requirements can include violation of security or other obligations related to Supply Chain Attacks, which like with the UK, leaves this method of compliance out. 

Expropriation in Irish law is derived initially from the Constitution, specifically Article 43(2)(2). Like the Danish constitution, the Common Good is the central point, as is ``occasion requires''. The latter refers to when the State can expropriate private property, which is the core protection of Article 43 outright. Land Laws\footnote{See, e.g., the Land and Conveyancing Law Reform Act 2009.} implement those powers for relevant situations, but like Denmark, there is theoretical room for potential expropriation of companies, although it seems unlikely for the same reasons as above.

\subsection{Regulating Adversarial Supply Chain Attacks in the Future}

In this section, we go through two potential future scenarios that may or may not justify the increased focus on mitigating Supply Chain Attacks, as well as some general thoughts on future legal mitigation approaches. 


\textit{Leveson, Nancy G.} once commented on a crucial assertion regarding the use of computers in general \parencite[P. 405]{leveson_safeware_1995}. \textit{"There is no technological imperative that says that we must use computers to control hazardous functions."} Increased uptake of technology that is vulnerable to certain types of adversarial attacks will result in increased successful attacks. It may be as simple as concluding that. But this kind of argumentation is pointless because it does not attempt to \textit{ex ante} predict and/or mitigate the failures. We know that increased automation may not result increased productivity, and that it may decrease safety of the system, and from what he have discussed earlier, it is also clear that it will decrease security. But, security can be improved, and this is where the discussion becomes more concrete. To show this, we take a look at a type of Supply Chain Attack that may become prevalent in the future, and which has worldwide consequences when it happens.

Shipping goods on ships is done with greatest profit and lowest cost per ton possible \cite{tolofari_shipping_1986}. If this can further be reduced, through automation and use of increased IoT, it is likely that the companies will make use of it. Furthermore, all ships of this caliber are tracked by the Automatic Identification System on a global scale, make use of GPS, make use of radars and if automated, would make use of a huge amount of new sensors and potentially robots, with no or few humans on deck. All of these subsystems/``subcontractors'' can be compromised, either individually or from the control systems suffering failures. The latter could be on the ship, or in the headquarters if they have a constant connection. As of the time of writing, there are measures in place to mitigate current attacks, ships can be sailed without any of these systems. But in the future, this may not be the case, and the entire infrastructure of the whole world may be at risk from Supply Chain Attacks \parencite{svilicic_paperless_2020}, and past incidents further support this \cite{meland_retrospective_2021}.

However, there is an even more pressing type of attack that can hit the very origin of CPS or IoT. Semiconductors, used to make processing and other power for the very devices that can be attacked, can be equally hit by Supply Chain Attacks \cite{voas_scarcity_2021}. Because there are extremely few main providers of these, the entire supply of the very basics of our digital infrastructure can be shut down in a matter of days. And while these attacks can hit the practical system of manufacturing or distributing them, the chips and other devices themselves can be attacked directly at the plant where they are produced, potentially compromising any computer or device they are part of \cite{dong_hardware_2020}. Together with hitting the supply chains through every transport type imaginable, the entire world economy is potentially at risk from Supply Chain Attacks in the future. While it is not on the lips of everyone right now, it may need to be in the future, and it will in the very least serve as justification for tighter and more nationally controlled standards, and perhaps technology specific requirements in law as well. 

\subsubsection{Future Regulatory Mitigation Techniques}

As we show in this Section, national jurisdictions do not always have many choices to prevent or otherwise regulate Supply Chain Attacks, which is a shame considering the potential consequences they can have. Extreme measures will rarely be used, and in a European context, regulation from the European Union is the best bet at horizontal hard legal rules to mitigate devastating attacks.

Otherwise, fines and very theoretical approaches to expropriation and emergency measures\footnote{During a state of emergency, many states can employ special written or unwritten rules beyond what is mentioned here, but these are so rarely seen and unclear, that we have not included them.} are not enough to fully and truly mitigate the attacks going forward. Technological developments are ongoing, but this does not mean that we need to throw the baby out with the bathwater. Existing enforcement measures in other areas can be reused, but a range of newer and more experimental enforcement measures could be considered, like financial incentives, tax breaks\footnote{There is a greater discussion on whether these or financial incentives have their intended effect as well.} or direct ministerial or public oversight. 






\subsection{Role of Guidance}

In all three jurisdictions and in the EU, guidance plays a huge or central role. It can be seen as attempting to bridge the gap between law and technology, and its existence is often justified by great technical or societal changes in contrast with the sometimes static hold of law. By its very nature, guidance is not \textit{per se} legally binding, and from the perspective of jurisprudence this is certainly the case. But law is about the perception of justice and both the individual or companies' perspective of the law \cite{peat_perception_2022}. To many, guidance and other guidelines or decisions, which are not legally binding, may feel or be perceived as such. This could be for professional reasons, there exists types of software engineers or managers who want to follow all ethically, as they perceive it, correct guidelines. This would clearly be for personal fulfillment, but the same could be said of the opposite kind, through those that would rather not follow any guidance other than internal and other employment orders or rules. This is exactly why the legal requirements in general are given on a cooperate basis, but this still leaves it for the individuals to decide on how to apply them in the end. In that sense, guidance acts as a nudging tool, or a legal tool enforceable by derivatives or mentions in statutes. The issue then becomes whether guidance becomes law. Anything that is clearly legally enforceable, even if not strictly a piece of legislation, can within certain legal philosophical approaches be considered a legal rule. Any kind of guidance that has this status will therefore be legal guidance. States and the EU should consider making these situations and choices clearer, instead of leaving it ambiguous, as this would include those are not willing to uphold non-legal rules on an individual plane, and ensure further compliance on a company level.


\subsection{Limits of Fines}


A core concept of criminology and the studies of punishment and enforcement, is that fines have very tight limits both in practice as well as economically\footnote{See a broader but worrying discussion of this in \parencite{omalley_theorizing_2009}.}. Punishing providers of security or others that can threaten the integrity of a supply chain or potentially threaten the digital infrastructure of a whole country or massive business conglomerates purely with fines, seems like a massive understatement of the potential consequences of such a neglect. Fines can potentially have other effects, such as loss of reputation for the individual or company that gets them issued, but the range of this kind of damage is disputed and likely very limited. 

From this, the choice of merely fining the providers and not employing stricter punishments in the form of threats of forceful closure or jail time for Directors or other responsible Officers, seems unwise going forward. 


\section{Future Work}

In-depth legal analysis of either how supply chains and critical infrastructure are regulated, or should be, is of high importance to follow this work, as would papers which explore organisational and other means to make these supply chains safer. Perhaps even from a classic safety engineering angle.

\section{Concluding Remarks} 

As we show above, Supply Chain Attacks are considered in various ways, but the means which we enforce compliance are not anywhere near enough to truly give incentive for all means of mitigation or possibly prevention possible. To reach this, we must regulate companies differently, or at the very minimum include the digital service providers or similar in our legal definitions of critical infrastructure, as they have already become part of it in practice\footnote{To this, see additional legal analysis by \parencite{massacci_economic_2016}.}. Furthermore, the lack of specialised rules in our examples should create uncertainty as to whether there will be appropriate sanctions in the future, to deal with the massive consequences preventable disruptions of supply chains may have on entire markets or at the very least countries. In this sense, digital and physical supply chains cannot exist without the companies, and giving them the same considerations as the shipping, train and flight companies which maintain the physical parts of supply chains would be wise. Finally, we would like to point out that guidance as a legal tool should be increasingly used, as it is our best shot at closing the gap between law and technology, but this increased usage should be done with more clarity, not less, to prevent confusion and possible circumvention.

\section{Funding}
The authors are supported by EPSRC funding under the PETRAS ROAST project. 

\printbibliography

@article{xenofontos_consumer_2022,
	title = {Consumer, {Commercial}, and {Industrial} {IoT} ({In}){Security}: {Attack} {Taxonomy} and {Case} {Studies}},
	volume = {9},
	issn = {23274662},
	doi = {10.1109/JIOT.2021.3079916},
	number = {1},
	journal = {IEEE Internet of Things Journal},
	author = {Xenofontos, Christos and Zografopoulos, Ioannis and Konstantinou, Charalambos and Jolfaei, Alireza and Khan, Muhammad Khurram and Choo, Kim Kwang Raymond},
	year = {2022},
	note = {Publisher: IEEE
\_eprint: 2105.06612},
	pages = {199--221},
}

@book{coleman_coding_2013,
	title = {Coding {Freedom}: {The} {Ethics} and {Aesthetics} of {Hacking}},
	publisher = {Princeton University Press},
	author = {Coleman, E. Gabriella},
	year = {2013},
	doi = {https://doi.org/10.1515/9781400845293},
}

@book{leveson_safeware_1995,
	edition = {1.},
	title = {Safeware: {System} {Safety} and {Computers}},
	isbn = {0-201-11972-2},
	publisher = {Addison-Wesley Publishing Company, Inc.},
	author = {Leveson, Nancy G.},
	year = {1995},
}

@techreport{miller_supply_2013,
	title = {Supply {Chain} {Attack} {Framework} and {Attack} {Patterns}},
	url = {https://apps.dtic.mil/sti/pdfs/ADA610495.pdf},
	number = {December 2013},
	institution = {MITRE},
	author = {Miller, John F},
	year = {2013},
	pages = {86},
}

@techreport{reed_supply_2014,
	title = {Supply {Chain} {Attack} {Patterns} : {Framework} and {Catalog}},
	url = {https://citeseerx.ist.psu.edu/viewdoc/download?doi=10.1.1.648.6043&rep=rep1&type=pdf},
	institution = {OFFICE OF THE DEPUTY ASSISTANT SECRETARY OF DEFENSE FOR SYSTEMS ENGINEERING},
	author = {Reed, Melinda and Miller, John F and Popick, Paul},
	year = {2014},
	pages = {88},
}

@article{eggers_nuclear_2020,
	title = {The nuclear digital {I}\&{C} system supply chain cyber-attack surface},
	volume = {122},
	issn = {0003018X},
	doi = {10.13182/T122-32483},
	number = {June},
	journal = {Transactions of the American Nuclear Society},
	author = {Eggers, Shannon L.},
	year = {2020},
	pages = {119--122},
}

@article{eggers_novel_2021,
	title = {A novel approach for analyzing the nuclear supply chain cyber-attack surface},
	volume = {53},
	issn = {2234358X},
	url = {https://doi.org/10.1016/j.net.2020.08.021},
	doi = {10.1016/j.net.2020.08.021},
	number = {3},
	journal = {Nuclear Engineering and Technology},
	author = {Eggers, Shannon},
	year = {2021},
	note = {Publisher: Elsevier Ltd},
	pages = {879--887},
}

@techreport{charney_cyber_2011,
	title = {Cyber {Supply} {Chain} {Risk} {Management}:{Toward} a {Global} {Vision} of {Transparency} and {Trust}},
	url = {http://download.microsoft.com/download/3/8/4/384483BA-B7B3-4F2F-9366-E83E4C7562D6/Cyber Supply Chain Risk Management white paper.pdf},
	institution = {Microsoft},
	author = {Charney, Scott and Werner, Eric T},
	year = {2011},
	pages = {19},
}

@article{kobara_cyber_2016,
	title = {Cyber physical security for {Industrial} {Control} {Systems} and {IoT}},
	volume = {E99D},
	issn = {17451361},
	doi = {10.1587/transinf.2015ICI0001},
	number = {4},
	journal = {IEICE Transactions on Information and Systems},
	author = {Kobara, Kazukuni},
	year = {2016},
	pages = {787--795},
}

@techreport{nsf_cyber-physical_2014,
	title = {Cyber-{Physical} {Systems}},
	institution = {National Science Foundation},
	author = {{NSF}},
	year = {2014},
	pages = {1--20},
}

@article{bonaci_make_2015,
	title = {To {Make} a {Robot} {Secure}: {An} {Experimental} {Analysis} of {Cyber} {Security} {Threats} {Against} {Teleoperated} {Surgical} {Robots}},
	url = {http://arxiv.org/abs/1504.04339},
	author = {Bonaci, Tamara and Herron, Jeffrey and Yusuf, Tariq and Yan, Junjie and Kohno, Tadayoshi and Chizeck, Howard Jay},
	year = {2015},
	note = {\_eprint: 1504.04339},
	pages = {1--11},
}

@article{roy_survey_2022,
	title = {Survey and {Taxonomy} of {Adversarial} {Reconnaissance} {Techniques}},
	issn = {0360-0300, 1557-7341},
	url = {https://dl.acm.org/doi/10.1145/3538704},
	doi = {10.1145/3538704},
	language = {en},
	journal = {ACM Computing Surveys},
	author = {Roy, Shanto and Sharmin, Nazia and Acosta, Jaime C. and Kiekintveld, Christopher and Laszka, Aron},
	month = may,
	year = {2022},
	pages = {3538704},
}

@article{zheng_interdiction_2019,
	title = {Interdiction models for delaying adversarial attacks against critical information technology infrastructure},
	volume = {66},
	issn = {0894-069X, 1520-6750},
	url = {https://onlinelibrary.wiley.com/doi/10.1002/nav.21859},
	doi = {10.1002/nav.21859},
	language = {en},
	number = {5},
	urldate = {2022-07-19},
	journal = {Naval Research Logistics (NRL)},
	author = {Zheng, Kaiyue and Albert, Laura A.},
	month = aug,
	year = {2019},
	pages = {411--429},
}

@article{zheng_robust_2019,
	title = {A {Robust} {Approach} for {Mitigating} {Risks} in {Cyber} {Supply} {Chains}},
	volume = {39},
	issn = {0272-4332, 1539-6924},
	url = {https://onlinelibrary.wiley.com/doi/10.1111/risa.13269},
	doi = {10.1111/risa.13269},
	language = {en},
	number = {9},
	urldate = {2022-07-19},
	journal = {Risk Analysis},
	author = {Zheng, Kaiyue and Albert, Laura A.},
	month = sep,
	year = {2019},
	pages = {2076--2092},
}

@article{zhou_impact_2022,
	title = {Impact of {Competition} from {Open} {Source} {Software} on {Proprietary} {Software}},
	volume = {31},
	issn = {1059-1478, 1937-5956},
	url = {https://onlinelibrary.wiley.com/doi/10.1111/poms.13575},
	doi = {10.1111/poms.13575},
	language = {en},
	number = {2},
	urldate = {2022-07-19},
	journal = {Production and Operations Management},
	author = {Zhou, Zach Zhizhong and Choudhary, Vidyanand},
	month = feb,
	year = {2022},
	pages = {731--742},
}

@incollection{hei_teleoperated_2022,
	title = {Teleoperated {Surgical} {Robot} {Security}: {Challenges} and {Solutions}},
	isbn = {978-1-79987-323-5 978-1-79987-325-9},
	shorttitle = {Teleoperated {Surgical} {Robot} {Security}},
	url = {http://services.igi-global.com/resolvedoi/resolve.aspx?doi=10.4018/978-1-7998-7323-5.ch009},
	language = {en},
	urldate = {2022-07-19},
	booktitle = {Advances in {Web} {Technologies} and {Engineering}},
	publisher = {IGI Global},
	author = {Al Momin, Md Abdullah and Islam, Md Nazmul},
	editor = {Hei, Xiali},
	year = {2022},
	doi = {10.4018/978-1-7998-7323-5.ch009},
	pages = {143--160},
}

@incollection{holt_social_2020,
	address = {Cham},
	title = {Social {Engineering}},
	isbn = {978-3-319-78439-7 978-3-319-78440-3},
	url = {http://link.springer.com/10.1007/978-3-319-78440-3_38},
	language = {en},
	urldate = {2022-07-19},
	booktitle = {The {Palgrave} {Handbook} of {International} {Cybercrime} and {Cyberdeviance}},
	publisher = {Springer International Publishing},
	author = {Bullée, Jan-Willem and Junger, Marianne},
	editor = {Holt, Thomas J. and Bossler, Adam M.},
	year = {2020},
	doi = {10.1007/978-3-319-78440-3_38},
	pages = {849--875},
}

@article{marelli_solarwinds_2022,
	title = {The {SolarWinds} hack: {Lessons} for international humanitarian organizations},
	volume = {104},
	issn = {1816-3831, 1607-5889},
	shorttitle = {The {SolarWinds} hack},
	url = {https://www.cambridge.org/core/product/identifier/S1816383122000194/type/journal_article},
	doi = {10.1017/S1816383122000194},
	language = {en},
	number = {919},
	urldate = {2022-07-19},
	journal = {International Review of the Red Cross},
	author = {Marelli, Massimo},
	month = apr,
	year = {2022},
	pages = {1267--1284},
}

@inproceedings{lagreca_survey_2017,
	title = {Survey on the {Insecurity} of the {Internet} of {Things}},
	language = {en},
	booktitle = {13th {International} {Workshop} on {Agents} and {Data} {Mining} {Interaction}},
	author = {LaGreca, Elizabeth and Boonthum-Denecke, Chutima},
	year = {2017},
	pages = {3},
}

@inproceedings{wang_insecurity_2021,
	address = {New Orleans Louisiana},
	title = {Insecurity of operational cellular {IoT} service: new vulnerabilities, attacks, and countermeasures},
	isbn = {978-1-4503-8342-4},
	shorttitle = {Insecurity of operational cellular {IoT} service},
	url = {https://dl.acm.org/doi/10.1145/3447993.3483239},
	doi = {10.1145/3447993.3483239},
	language = {en},
	urldate = {2022-07-19},
	booktitle = {Proceedings of the 27th {Annual} {International} {Conference} on {Mobile} {Computing} and {Networking}},
	publisher = {ACM},
	author = {Wang, Sihan and Tu, Guan-Hua and Lei, Xinyu and Xie, Tian and Li, Chi-Yu and Chou, Po-Yi and Hsieh, Fucheng and Hu, Yiwen and Xiao, Li and Peng, Chunyi},
	month = oct,
	year = {2021},
	pages = {437--450},
}

@article{yousefnezhad_security_2020,
	title = {Security in product lifecycle of {IoT} devices: {A} survey},
	volume = {171},
	issn = {10848045},
	shorttitle = {Security in product lifecycle of {IoT} devices},
	url = {https://linkinghub.elsevier.com/retrieve/pii/S1084804520302538},
	doi = {10.1016/j.jnca.2020.102779},
	language = {en},
	urldate = {2022-07-19},
	journal = {Journal of Network and Computer Applications},
	author = {Yousefnezhad, Narges and Malhi, Avleen and Främling, Kary},
	month = dec,
	year = {2020},
	pages = {102779},
}

@misc{kamara_misaligned_2021,
	title = {Misaligned {Union} laws? {A} comparative analysis of certification in the {Cybersecurity} {Act} and the {General} {Data} {Protection} {Regulation}},
	language = {en},
	author = {Kamara, Irene},
	year = {2021},
}

@article{casarosa_cybersecurity_2022,
	title = {Cybersecurity certification of {Artificial} {Intelligence}: a missed opportunity to coordinate between the {Artificial} {Intelligence} {Act} and the {Cybersecurity} {Act}},
	volume = {3},
	issn = {2662-9720, 2662-9739},
	shorttitle = {Cybersecurity certification of {Artificial} {Intelligence}},
	url = {https://link.springer.com/10.1365/s43439-021-00043-6},
	doi = {10.1365/s43439-021-00043-6},
	language = {en},
	number = {1},
	urldate = {2022-07-19},
	journal = {International Cybersecurity Law Review},
	author = {Casarosa, Federica},
	month = jun,
	year = {2022},
	pages = {115--130},
}

@article{omalley_theorizing_2009,
	title = {Theorizing fines},
	volume = {11},
	issn = {1462-4745, 1741-3095},
	url = {http://journals.sagepub.com/doi/10.1177/1462474508098133},
	doi = {10.1177/1462474508098133},
	language = {en},
	number = {1},
	urldate = {2022-07-19},
	journal = {Punishment \& Society},
	author = {O'Malley, Pat},
	month = jan,
	year = {2009},
	pages = {67--83},
}

@article{svilicic_paperless_2020,
	title = {Paperless ship navigation: cyber security weaknesses},
	volume = {13},
	issn = {1938-7741, 1938-775X},
	shorttitle = {Paperless ship navigation},
	url = {https://link.springer.com/10.1007/s12198-020-00222-2},
	doi = {10.1007/s12198-020-00222-2},
	language = {en},
	number = {3-4},
	urldate = {2022-07-20},
	journal = {Journal of Transportation Security},
	author = {Svilicic, Boris and Kristić, Miho and Žuškin, Srđan and Brčić, David},
	month = dec,
	year = {2020},
	pages = {203--214},
}

@article{creazza_who_2022,
	title = {Who cares? {Supply} chain managers’ perceptions regarding cyber supply chain risk management in the digital transformation era},
	volume = {27},
	language = {en},
	number = {1},
	journal = {Supply Chain Management},
	author = {Creazza, Alessandro and Colicchia, Claudia and Spiezia, Salvatore and Dallari, Fabrizio},
	year = {2022},
	pages = {24},
}

@article{ameen_keeping_2021,
	title = {Keeping customers' data secure: {A} cross-cultural study of cybersecurity compliance among the {Gen}-{Mobile} workforce},
	volume = {114},
	issn = {07475632},
	shorttitle = {Keeping customers' data secure},
	url = {https://linkinghub.elsevier.com/retrieve/pii/S0747563220302831},
	doi = {10.1016/j.chb.2020.106531},
	language = {en},
	urldate = {2022-07-24},
	journal = {Computers in Human Behavior},
	author = {Ameen, Nisreen and Tarhini, Ali and Shah, Mahmood Hussain and Madichie, Nnamdi and Paul, Justin and Choudrie, Jyoti},
	month = jan,
	year = {2021},
	pages = {106531},
}

@article{carr_uk_2018,
	title = {{UK} cybersecurity industrial policy: an analysis of drivers, market failures and interventions},
	volume = {3},
	issn = {2373-8871, 2373-8898},
	shorttitle = {{UK} cybersecurity industrial policy},
	url = {https://www.tandfonline.com/doi/full/10.1080/23738871.2018.1550523},
	doi = {10.1080/23738871.2018.1550523},
	language = {en},
	number = {3},
	urldate = {2022-07-24},
	journal = {Journal of Cyber Policy},
	author = {Carr, Madeline and Tanczer, Leonie Maria},
	month = sep,
	year = {2018},
	pages = {430--444},
}

@article{kilovaty_privatized_2019,
	title = {Privatized {Cybersecurity} {Law}},
	issn = {1556-5068},
	url = {https://www.ssrn.com/abstract=3338155},
	doi = {10.2139/ssrn.3338155},
	language = {en},
	urldate = {2022-07-24},
	journal = {SSRN Electronic Journal},
	author = {Kilovaty, Ido},
	year = {2019},
}

@article{boeke_national_2018,
	title = {National cyber crisis management: {Different} {European} approaches},
	volume = {31},
	issn = {09521895},
	shorttitle = {National cyber crisis management},
	url = {https://onlinelibrary.wiley.com/doi/10.1111/gove.12309},
	doi = {10.1111/gove.12309},
	language = {en},
	number = {3},
	urldate = {2022-07-24},
	journal = {Governance},
	author = {Boeke, Sergei},
	month = jul,
	year = {2018},
	pages = {449--464},
}

@article{nguyen_industrial_2019,
	title = {Industrial {Internet} of {Things}, {Big} {Data}, and {Artificial} {Intelligence} in the {Smart} {Factory}: a survey and perspective},
	language = {en},
	author = {Nguyen, H D and Tran, K P and Zeng, X and Koehl, L and Castagliola, P and Bruniaux, Pascal},
	year = {2019},
	pages = {6},
}

@article{tucker_sustainable_2021,
	title = {Sustainable {Product} {Lifecycle} {Management}, {Industrial} {Big} {Data}, and {Internet} of {Things} {Sensing} {Networks} in {Cyber}-{Physical} {System}-based {Smart} {Factories}},
	volume = {6},
	issn = {2329-4175},
	url = {https://addletonacademicpublishers.com/contents-jsme/2091-volume-9-1-2021/3944-sustainable-product-lifecycle-management-industrial-big-data-and-internet-of-things-sensing-networks-in-cyber-physical-system-based-smart-factories},
	doi = {10.22381/jsme9120211},
	language = {en},
	number = {1},
	urldate = {2022-07-25},
	journal = {Journal of Self-Governance and Management Economics},
	author = {Tucker, Glenn},
	year = {2021},
	pages = {9},
}

@inproceedings{de_las_morenas_security_2020,
	address = {Buenos Aires, Argentina},
	title = {Security {Experiences} in {IoT} based applications for {Building} and {Factory} {Automation}},
	isbn = {978-1-72815-754-2},
	url = {https://ieeexplore.ieee.org/document/9067229/},
	doi = {10.1109/ICIT45562.2020.9067229},
	language = {en},
	urldate = {2022-07-25},
	booktitle = {2020 {IEEE} {International} {Conference} on {Industrial} {Technology} ({ICIT})},
	publisher = {IEEE},
	author = {de las Morenas, Javier and da Silva, Carolina Miller and Funchal, Gustavo Silva and Melo, Victoria and Vallim, Marcos and Leitao, Paulo},
	month = feb,
	year = {2020},
	pages = {322--327},
}

@article{scarman_human_2012,
	title = {Human {Rights} in an {Unwritten} {Constitution}},
	volume = {2},
	issn = {0269-1922},
	url = {http://www.ubplj.org/index.php/dlj/article/view/163},
	doi = {10.5750/dlj.v2i1.163},
	language = {en},
	number = {1},
	urldate = {2022-08-03},
	journal = {The Denning Law Journal},
	author = {Scarman, The Rt Hon Lord},
	month = oct,
	year = {2012},
	pages = {129--135},
}

@article{frosini_is_2019,
	title = {Is {Brexit} {Ripping} up the {Unwritten} {Constitution} of the {United} {Kingdom}?},
	volume = {11},
	copyright = {HeinOnline},
	number = {1},
	journal = {Italian Journal of Public Law},
	author = {Frosini, Justin O.},
	year = {2019},
}

@article{gyenes_voluntary_2014,
	title = {A {Voluntary} {Cybersecurity} {Framework} {Is} {Unworkable}- {Government} {Must} {Crack} the {Whip}},
	volume = {14},
	number = {2},
	urldate = {2022-08-03},
	journal = {Pittsburgh Journal of Technology Law and Policy},
	author = {Gyenes, Robert},
	year = {2014},
	pages = {293--314},
}

@article{shackelford_bottoms_2016,
	title = {{BOTTOMS} {UP}: {A} {COMPARISON} {OF} "{VOLUNTARY}" {CYBERSECURITY} {FRAMEWORKS}},
	volume = {16},
	language = {en},
	author = {Shackelford, J and Russell, Scott and Haut, Jeffrey},
	year = {2016},
	pages = {45},
}

@article{haber_cybersecurity_2017,
	title = {{CYBERSECURITY} {FOR} {INFRASTRUCTURE}: {A} {CRITICAL} {ANALYSIS}},
	volume = {44},
	number = {2},
	journal = {Florida State University Law Review},
	author = {Haber, Eldar and Zarsky, Tal},
	year = {2017},
}

@article{massacci_economic_2016,
	title = {Economic {Impacts} of {Rules}- versus {Risk}-{Based} {Cybersecurity} {Regulations} for {Critical} {Infrastructure} {Providers}},
	volume = {14},
	number = {3},
	urldate = {2022-08-04},
	journal = {IEEE Security \& Privacy},
	author = {Massacci, Fabio and Ruprai, Raminder and Collinson, Matthew and Williams, Julian},
	month = may,
	year = {2016},
	pages = {52--60},
}

@article{duan_monopolies_2020,
	title = {{OF} {MONOPOLIES} {AND} {MONOCULTURES}: {THE} {INTERSECTION} {OF} {PATENTS} {AND} {NATIONAL} {SECURITY}},
	volume = {36},
	language = {en},
	number = {4},
	journal = {Santa Clara High Technology Law Journal},
	author = {Duan, Charles},
	year = {2020},
	pages = {39},
}

@article{economides_restrictions_2019,
	title = {Restrictions on {Privacy} and {Exploitation} in the {Digital} {Economy}: {A} {Competition} {Law} {Perspective}},
	journal = {CLES Research Paper Series},
	author = {Economides, Nick and Lianos, Ioannis},
	year = {2019},
}

@article{pisaric_communications_2022,
	title = {Communications {Encryption} as an {Investigative} {Obstacle}},
	volume = {60},
	language = {en},
	number = {1},
	urldate = {2022-08-04},
	journal = {Journal of Criminology and Criminal Law},
	author = {Pisaric, Milana},
	year = {2022},
	pages = {61--74},
}

@article{anderson_review_2021,
	title = {Review of \textit{{Crypto} {Wars}—{The} {Fight} for {Privacy} in the {Digital} {Age}: {A} {Political} {History} of {Digital} {Encryption}}},
	issn = {0161-1194, 1558-1586},
	shorttitle = {Review of {\textless}i{\textgreater}{Crypto} {Wars}—{The} {Fight} for {Privacy} in the {Digital} {Age}},
	url = {https://www.tandfonline.com/doi/full/10.1080/01611194.2021.2002977},
	doi = {10.1080/01611194.2021.2002977},
	language = {en},
	urldate = {2022-08-04},
	journal = {Cryptologia},
	author = {Anderson, Patrick D.},
	month = dec,
	year = {2021},
	pages = {1--14},
}

@article{hellegren_history_2017,
	title = {A history of crypto-discourse: encryption as a site of struggles to define internet freedom},
	volume = {1},
	issn = {2470-1475, 2470-1483},
	shorttitle = {A history of crypto-discourse},
	url = {https://www.tandfonline.com/doi/full/10.1080/24701475.2017.1387466},
	doi = {10.1080/24701475.2017.1387466},
	language = {en},
	number = {4},
	urldate = {2022-08-04},
	journal = {Internet Histories},
	author = {Hellegren, Z. Isadora},
	month = sep,
	year = {2017},
	pages = {285--311},
}

@article{ellis_history_1999,
	title = {{THE} {HISTORY} {OF} {NON}-{SECRET} {ENCRYPTION}},
	volume = {23},
	issn = {0161-1194, 1558-1586},
	url = {http://www.tandfonline.com/doi/abs/10.1080/0161-119991887919},
	doi = {10.1080/0161-119991887919},
	language = {en},
	number = {3},
	urldate = {2022-08-04},
	journal = {Cryptologia},
	author = {Ellis, J. H.},
	month = jul,
	year = {1999},
	pages = {267--273},
}

@inproceedings{suryotrisongko_review_2019,
	address = {Kaohsiung, Taiwan},
	title = {Review of {Cybersecurity} {Research} {Topics}, {Taxonomy} and {Challenges}: {Interdisciplinary} {Perspective}},
	language = {en},
	urldate = {2022-08-04},
	publisher = {IEEE},
	author = {Suryotrisongko, Hatma and Musashi, Yasuo},
	month = nov,
	year = {2019},
	pages = {162--167},
}

@incollection{choo_interdisciplinary_2020,
	address = {Cham},
	title = {Interdisciplinary {Cybersecurity}: {Rethinking} the {Approach} and the {Process}},
	volume = {1055},
	isbn = {978-3-030-31238-1 978-3-030-31239-8},
	shorttitle = {Interdisciplinary {Cybersecurity}},
	url = {http://link.springer.com/10.1007/978-3-030-31239-8_6},
	language = {en},
	urldate = {2022-08-04},
	booktitle = {National {Cyber} {Summit} ({NCS}) {Research} {Track}},
	publisher = {Springer International Publishing},
	author = {Jacob, Johanna and Peters, Michelle and Yang, T. Andrew},
	editor = {Choo, Kim-Kwang Raymond and Morris, Thomas H. and Peterson, Gilbert L.},
	year = {2020},
	doi = {10.1007/978-3-030-31239-8_6},
	note = {Series Title: Advances in Intelligent Systems and Computing},
	pages = {61--74},
}

@article{tolofari_shipping_1986,
	title = {Shipping {Costs} and the {Controversy} {Over} {Open} {Registry}},
	volume = {34},
	issn = {00221821},
	url = {https://www.jstor.org/stable/2098626?origin=crossref},
	doi = {10.2307/2098626},
	language = {en},
	number = {4},
	urldate = {2022-08-05},
	journal = {The Journal of Industrial Economics},
	author = {Tolofari, S. R. and Button, K. J. and Pitfield, D. E.},
	month = jun,
	year = {1986},
	pages = {409},
}

@article{meland_retrospective_2021,
	title = {A {Retrospective} {Analysis} of {Maritime} {Cyber} {Security} {Incidents}},
	volume = {15},
	issn = {2083-6473},
	url = {http://www.transnav.eu/Article_A_Retrospective_Analysis_of_Maritime_Cyber_Security_Incidents_Meland,59,1144.html},
	doi = {10.12716/1001.15.03.04},
	language = {en},
	number = {3},
	urldate = {2022-08-05},
	journal = {TransNav, the International Journal on Marine Navigation and Safety of Sea Transportation},
	author = {Meland, Per HÁkon and Bernsmed, Karin and Wille, Egil and Rødseth, Ørnulf Jan and Nesheim, Dag Atle},
	year = {2021},
	pages = {519--530},
}

@article{voas_scarcity_2021,
	title = {Scarcity and {Global} {Insecurity}: {The} {Semiconductor} {Shortage}},
	volume = {23},
	issn = {1520-9202, 1941-045X},
	shorttitle = {Scarcity and {Global} {Insecurity}},
	url = {https://ieeexplore.ieee.org/document/9568259/},
	doi = {10.1109/MITP.2021.3105248},
	language = {en},
	number = {5},
	urldate = {2022-08-05},
	journal = {IT Professional},
	author = {Voas, Jeffrey and Kshetri, Nir and DeFranco, Joanna F.},
	month = sep,
	year = {2021},
	pages = {78--82},
}

@article{dong_hardware_2020,
	title = {Hardware {Trojans} in {Chips}: {A} {Survey} for {Detection} and {Prevention}},
	volume = {20},
	issn = {1424-8220},
	shorttitle = {Hardware {Trojans} in {Chips}},
	url = {https://www.mdpi.com/1424-8220/20/18/5165},
	doi = {10.3390/s20185165},
	language = {en},
	number = {18},
	urldate = {2022-08-05},
	journal = {Sensors},
	author = {Dong, Chen and Xu, Yi and Liu, Ximeng and Zhang, Fan and He, Guorong and Chen, Yuzhong},
	month = sep,
	year = {2020},
	pages = {5165},
}

@article{peat_perception_2022,
	title = {Perception and {Process}: {Towards} a {Behavioural} {Theory} of {Compliance}},
	volume = {13},
	issn = {2040-3593},
	shorttitle = {Perception and {Process}},
	url = {https://academic.oup.com/jids/article/13/2/179/6439208},
	doi = {10.1093/jnlids/idab030},
	language = {en},
	number = {2},
	urldate = {2022-08-05},
	journal = {Journal of International Dispute Settlement},
	author = {Peat, Daniel},
	month = jun,
	year = {2022},
	pages = {179--209},
}

\end{document}